# Theorizing neuro-induced relationships between cognitive diversity, motivation, grit and academic performance in multidisciplinary engineering education context

Duy Duong-Tran[1,2], Siqing Wei[3], Li Shen[1]

1 Department of Biostatistics, Epidemiology and Informatics, Perelman School of Medicine at the University of Pennsylvania, Philadelphia, Pennsylvania
2 Department of Mathematics, United States Naval Academy, Annapolis, Maryland
3 School of Engineering Education, Purdue University, West Lafayette, Indiana

**Abstract**
Nowadays, engineers need to tackle many unprecedented challenges that are often complex, and, most importantly, cannot be exhaustively compartmentalized into a single engineering discipline. In other words, most engineering problems need to be solved from a multidisciplinary approach. However, conventional engineering programs usually adopt pedagogical approaches specifically tailored to traditional, niched engineering disciplines, which become increasingly deviated from the industry needs as those programs are typically designed and taught by instructors with highly specialized engineering training and credentials. To reduce the gap, more multidisciplinary engineering programs emerge by systematically stretching across all engineering fibers, and challenge the sub-optimal traditional pedagogy crowded in engineering classrooms. To further advance future-oriented pedagogy, in this work, we hypothesized neuro-induced linkages between how cognitively different learners are and how the linkages would affect learners in the knowledge acquisition process. We situate the neuro-induced linkages in the context of multidisciplinary engineering education and propose possible pedagogical approaches to actualize the implications of this conceptual framework. Our study, based on the innovative concept of brain fingerprint, would serve as a pioneer model to theorize key components of learner-centered multidisciplinary engineering pedagogy which centers on the key question: how do we motivate engineering students of different backgrounds from a neuro-inspired perspective?

**Introduction**
In today's society, the majority of encountered challenges are complex, open-ended, and hard to effectively categorize into any single particular discipline [1], which requires cross-field collaborations. To effectively facilitate both multi- and intra-/trans disciplinary collaboration, engineers need to access and understand the grounded concepts from other fields to tackle complex challenges. For instance, construction engineers are now facing the challenge of highway, railroad, and infrastructure design in response to climate change risks. As such, they need to adopt new approaches in design to accommodate flooding risk or increased temperature changes and usually consult mechanical engineers during the innovation and implementation phases for such enhancement [2]. Furthermore, from the environmental perspective to enhance sustainability and life cycle, construction engineers need to, at least, acquire an understanding of concepts and potential impact of their design of climate change from environmental engineers.

Similarly, the emergence of network neuroscience (or brain connectomics) is attributed to interdisciplinary efforts [3]. The birth of network neuroscience stemmed from the surge of social networks in the social science field inspired by the applied mathematics field of "network science", which used the mathematical language of networks (graphs) to analyze patterns and dynamics of social connections. Parallelly, the last decade has witnessed the growth of technologies of magnetic resonance imaging (MRI) which presented the world with high-quality data on human organisms (e.g., human brain imaging). Such developments have led to an organic fusion of network science and neuroscience that yielded the emerging interdisciplinary field of network neuroscience. Since its birth, network neuroscience has absorbed contributions from many disciplines, including but not limited to psychology, neuroscience, mathematics and physics, and produced many clinically oriented breakthroughs, such as clinical utility of brain connectomics in the context of brain fingerprint [3, 4, 5, 6, 7, 8, 9, 10].

In both engineering examples, we could infer that current challenges cannot be solved within a single discipline and usually need broader consideration of society and humanity beyond purely technical solutions [2, 11]. Consequently, well-trained engineers are now required to either grasp knowledge of all kinds or, more viable, collaborate with others from different domains to tackle these challenges and provide viable solutions [12]. The cross-discipline collaborations can be categorized into three main types: multidisciplinary, transdisciplinary, and interdisciplinary problems. In this paper, we group interdisciplinary and transdisciplinary discipline (hereafter, denoted as Type I) under one category due to its closeness in definitions [2]; multidisciplinary discipline will be referred to as Type II. Specifically, multidisciplinary approach tailors to problems whose solutions can be analyzed from multiple perspectives (e.g., multiple disciplinary angles) and can be independently proposed, in part, by the relevant disciplines (e.g., A, B, and C in Figure 1, bottom left panel). One thing to note is that multidisciplinary collaborations do necessarily draw information and knowledge (hence, exchange of information) from different disciplines but do not require an integrated, fused approach [2]. On the other hand, both inter- and trans- disciplinary approaches require an integrated, fused approach that necessitates a shared, well-integrated solution that often yields an emergent interdisciplinary field of knowledge domain [2]. However, for this paper, we neglect the difference between inter- and trans- disciplinary approaches, which falls on whether to break down the disciplinary boundaries. However, this cross-discipline organic fusion of knowledge often is irreversible in the sense that they often produce new knowledge that needs independent documentation and a well-developed faculty of understanding.

The most striking commonality between the two types is that they both require organic collaborations and employment of various ideas and perspectives from multiple domains [2]. On the other hand, the fundamental difference between them, in terms of collaboration, centers around their organization. Specifically, inter-/trans- disciplinary relies on integrating, fusing, and sometimes irreversible merging of knowledge and expertise while multi-disciplinary often forms de-compartmentalized and reversible collaborations. As such, inter-/trans-disciplinary solution space often emerges from a newly-developed field that grows out of two or more well-defined disciplines (e.g., field [ABC] Figure 1, top right panel).

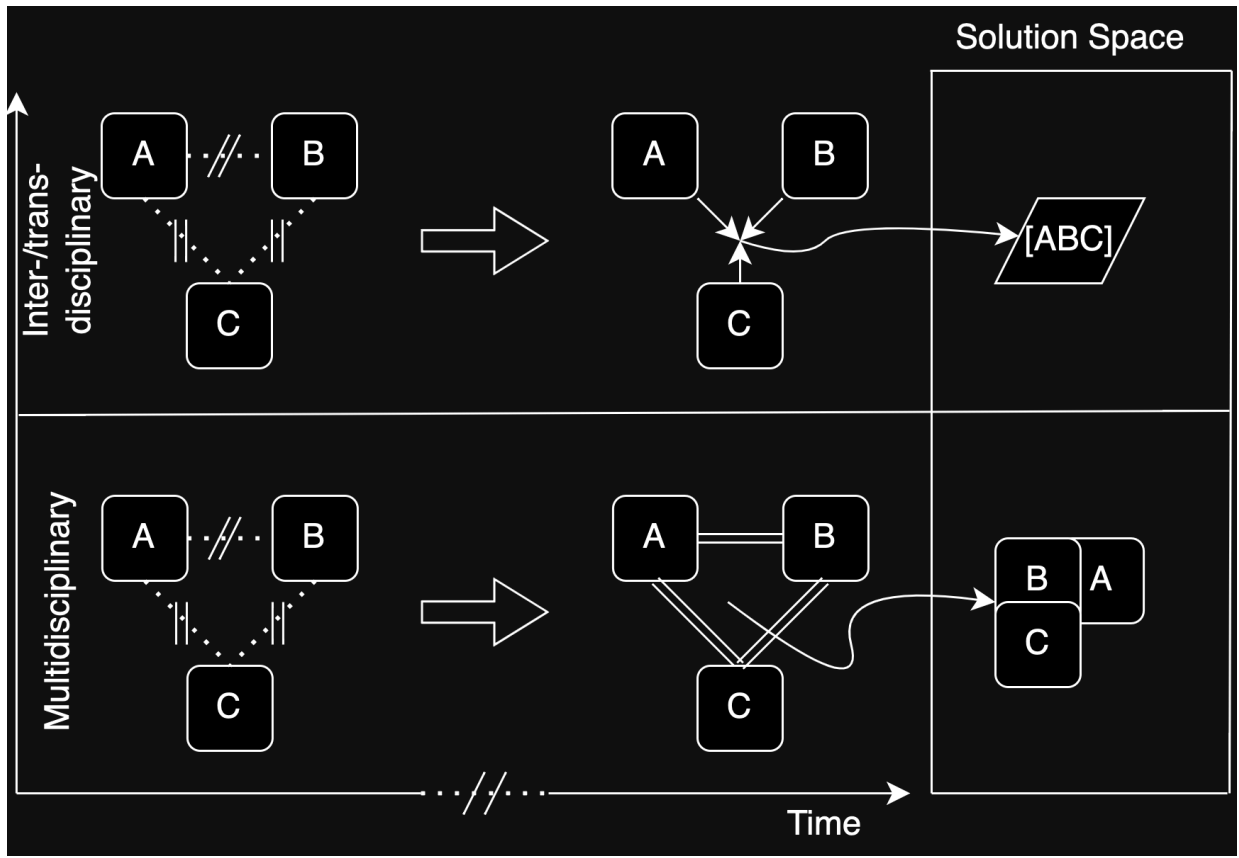

**Figure 1: A schematic representation of inter-/trans- (Type I) and multi- (Type II) disciplinary engineering in the domain of time**. Both disciplines yield collaborations from at least two independent fields. Of note, in the top panel, interdisciplinary collaboration results in an emergent field [ABC] that requires a complete rethinking and development from interdisciplinary fields A, B, and C. In the bottom panel, multidisciplinary collaboration, over time, might bring A, B, and C disciplines "closer" but does not result in an emergent discipline. Note that //'s on the dashed lines denote the independence between the disciplines while the solid ||'s represents the existence of commonalities between disciplines.

Over time, these organic fusions induced by inter-/trans-disciplinary approaches cannot be effectively and exhaustively categorized into any single, isolated, independent mother fields (e.g., squares A, B or C Figure 1, top left panel). The field of interdisciplinary engineering (Figure 1, top panel) is created from the necessity to provide structured methods of education that can effectively bring together the "*combinations of theories, concepts and methods from different disciplines in a single context*" [13]. In other words, inter-/trans- disciplinary engineering research focuses on the integration of concepts, theories, data and expertise [2, 14]. On the other hand, multi-disciplinary collaborations (Figure 1, bottom panel) are those that also emerge from more than two disciplines but these challenges can be effectively compartmentalized into isolated sub-problems that can be fully characterized under correspondingly independent disciplines or can be looked at from multiple disciplinary angles [2].

Along with the surge of increasingly complex engineering challenges, recently, network neuroscientists have discovered that each individual possesses a unique cognitive signature of information processing, reasoning, and learning [15, 16]. This concept is denominated as brain fingerprint (equivalently, cognitive diversity). Collectively, the emergent concept of brain fingerprint sheds light on consequential discoveries on potential mechanisms of learning trajectories among the human population. In the context of engineering education, these findings would, in turn, provide viable tools to infer latent links between the cognitively diverse sources of motivation, academic performance, and engineering major selection.

Collectively, the surge of multidisciplinary challenges, coupled with increasing evidence of cognitive diversity, not only increases the societal expectations for in-training engineers to supply viable, long-lasting solutions [2], but also urges a comprehensive, proactive approach to rethink engineering education in the context of multi-/inter-/trans- disciplinary institutionalizing process. This paper aims to theorize the relationship between cognitive diversity, motivation, and academic performance in the context of multidisciplinary or first-year engineering coursework that delivers a knowledge domain spanned across multiple engineering disciplines. The rest of the paper is structured as follows: Section 2 provides a brief literature review of key components of this conceptual framework; Section 3 argues the relationship among these components; and finally, Section 4 proposes some possible interpretations based on the theorized relationships.

**Section 2 Literature review**
This section revisits and summarizes student major choice and academic retention from the perspectives of neuroscience, specifically focusing on intrinsic motivation. The aim is to build a foundational support to theorize the relationship among them in Section 3.

**2.1. An Overview of Educational Neuroscience:**
The overarching goal of educational neuroscience is to provide neuro-inspired strategies for teaching. As such, educational neuroscience is an emerging interdisciplinary research field that brings together expertise in neuroscience and cognitive science to the realm of education [17].

Recent advancements in neuroscientific technology, especially Magnetic Resonance Imaging (MRI), have enabled researchers in this field to induce many valuable insights into the learning mechanisms of the human brain [15], which opens a new door to investigate learning science and associated pedagogy. Being an emerging field, education neuroscience faces unavoidable challenges [18] to legitimize and position itself, relative to neuroscience and education. In this paper, we argue that neuroscience research, if leveraged correctly, could prove to uniquely transform teaching practices. We construct our argument in the context of motivation/grit and their association with cognitive and neuro diversity. It is worthy to note that our theoreticized relationships among these cognitive processes could be applied beyond this paper focus: multidisciplinary engineering coursework.

## 2.2. Brain Fingerprint, cognitive diversity in the context of educational neuroscience:

The discovery of unique patterns of ridges and sweat glands on fingertips by physician Marcello Malpighi triggered a major breakthrough, in terms of scientific methods to uniquely identify individuals based on fingerprints. Fasting forward into the 21st century, the fingerprint concept has now extended to other bio-data sources (e.g., voice recognition and MRI). This unprecedented development has enabled neuroimaging researchers to investigate the individual-specific level of functional and structural connections in the human brain, modeled as networks. Structural physical white-matter connections between gray matter brain regions of interest are typically quantified from diffusion-weighted imaging (DWI) data. Functional connections are modeled from functional magnetic resonance imaging (FMRI) data, by measuring temporal statistical dependencies between brain region pairs. According to Chrysochoou and colleagues [19], neurodiversity refers to neurological differences leading to "distinct cognitive characteristics" that dictate learning differences among individuals. In this paper, neurodiversity was defined without proper background of neuroscience. As such, the term neurodiversity is not typically utilized in the context of basic science (e.g., brain fingerprint in the domain of brain connectivity) but rather in the context of neurological disorders (e.g., ADHD, autism, or dyslexia). Specifically, the author proposed the term neurodiversity without demonstrating whether the degree of "fingerprint" (individuals' cognitive signatures) would decrease or increase in neurodegenerative groups such as ADHD, autism, compared to healthy controls. With recent advancement in system neuroscience, we can formally define how different individuals possess different cognitive signatures, especially when it comes to knowledge acquisition, through the concept and study of brain fingerprinting. In our paper, we propose and implement the concept of brain fingerprint into the context of multidisciplinary engineering education.

## 2.3. Motivation

One of the most critical components in our paper is centered around motivation, which is a critical construct related to academic achievement [20, 21]. In the context of engineering education, motivation plays a critical role in shaping students' success, manifested compartmentally in individual coursework and collectively in the entire program [22]. Motivation can be classified into two types: intrinsic and extrinsic. In this paper, we focus on linking motivation in the context of self-determination which was proposed by Ryan and Deci [23, 24] In particular, Ryan and Deci concluded that "*both intrinsic motivation and well-internalized forms of extrinsic motivation predict an array of positive outcomes*". In their earlier publication [23], the authors also contended that "*people can be motivated because they value an activity or because there is strong external coercion… abiding interest or a bribe… commitment to excel or fear of being surveilled.*" Meanwhile, the authors argued that "*placing strong relative importance on intrinsic aspiration was positively associated with well-being indicators such as self-esteem, self-actualization, the inverse of depression and anxiety.*" Further, a more autonomous form of teaching would lead to a higher level of engagement [23]. Evidently, there exists much more depth and perhaps, blind spots when it comes to how to improve the overall students' experience with a particular coursework, rather than the assumption that "they (the students) take the course, they must be motivated" and that it is not our (the instructors') jobs to motivate them to learn. Other studies such as Savage and colleagues [25] also suggested that extrinsic motivation does not provide and promote healthy, long-lasting forms of teaching practices. In section 3, we will theorize and induce the relationship between cognitive diversity and motivation, especially intrinsic motivation.

**2.4. Grit and motivation**

Grit, as proposed by Duckworth and colleagues [26], is defined as both perseverance and passion for long-term goals, which are the key component for academic retention and performance. Indeed, grit was shown to be the key indicator and a strong predictor of service academies. Grit's key component is the perseverance of effort, driven by consistency of interest, and can be measured by Grit-Scale (Grit-S) [27]. In this context, grit has a very intimate relationship with intrinsic motivation. Specifically, in the aforementioned papers, the authors argued that grit was a better predictor of long-term retention and success, compared to other indicators such as grade point average (GPA). The key take-away in Duckworth and colleagues' paper [26] was that i) grit is an individual-driven factor in a professional's goal achievement and ii) grit demonstrated a "*incremental predictive validity*" of success. For a comprehensive review of grit's role in engineering education, please refer to the work of Direito and colleagues in 2021 [26]. In this paper, our focus is on theorizing the relationship between grit, motivation with academic performance using Duckworth's viewpoint. As such, we ground Duckworth's work on grit and argue that motivation, grit, and academic performance not only have causal relationships but also are highly individually driven, which is related to brain fingerprint and cognitive diversity.

**Section 3. The formal construction of a neuro-inspired conceptual framework**

In this section, we aim to describe the relationships among the components introduced in Section 2: cognitive diversity, intrinsic motivation, grit, and college major selection. Specifically, we propose a model to link these aforementioned components and describe these associations. Here, the directed arrows mean causal relationships (e.g., A causes B) and undirected links represent correlation (e.g., A is related to B, and vice versa). Further, the thickness of the arrows represents the relative weights of either causal link or correlating links in the model.

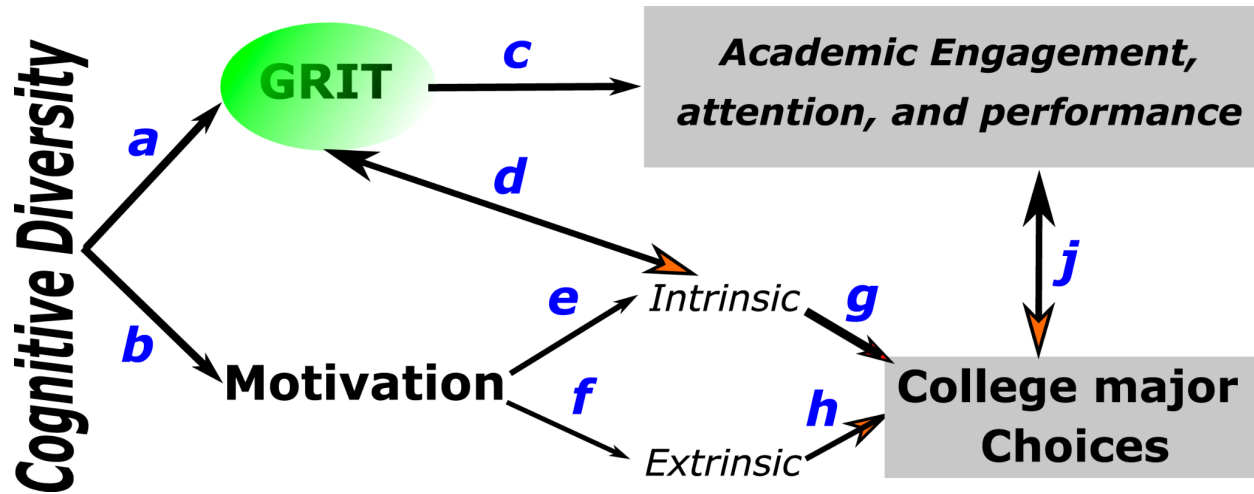

**Figure 2. A schematic diagram of the theorizing relationships among cognitive diversity, grit, motivation, academic performance/engagement/attention, and major choices**

### 3.1. Relationship between cognitive diversity and motivation, grit (link a, b):

In Section 2., we hypothesize that there is a causal relationship between brain fingerprint, cognitive diversity and motivation (Section 2.3) and grit (Section 2.4.). In our model, we propose two hypothesized links (Figure 2, a, b), which indicate that cognitive diversity, as quantified by brain fingerprint concept [9], is the key driver for individualized grit and motivation. Specifically, individual professional achievement, according to [26], "*... less is known about other individual differences that predict success*." This means that grit is highly correlated with an individual's cognitive signatures that are driven by brain fingerprints. As such cognitive diversity is necessarily a causal factor that drives grit (e.g., directed link *a* in the model). Additionally, there are studies in system neuroscience which indicate a high degree of individuality when it comes to rewarding mechanisms of brain networks [9] (e.g., directed link *b* in the model). The reward mechanism in the human brain is not necessarily tied with either intrinsic or extrinsic motivation, exclusively, but rather stimulus dependent. In neuroscience studies, responses to rewards are usually external as internally motivated rewards are convoluted to simulate. In this paper, we theorize that cognitive diversity necessarily infers a causal link (link *b*) with both intrinsic and extrinsic motivation.

### 3.2. Relationship between grit and motivation (Link d, e, f):

Through the theorized link (Figure 2, **d**), based upon the discussions of Section 2.3 and 2.4, we theorized the relationship between grit (persistence, perseverance) and intrinsic motivation. Note that link **e, f** was established from prior studies (Section 2.3 and 2.4). In this section, we hypothesize that there exists an undirected link (e.g., link **d**) between grit and intrinsic motivation. In [26], the authors pointed out that grit's "under-the-hood" engine is perseverance which has been linked with long-term, consistent source of intrinsic motivation, as extrinsic motivation is typically episodic and inconsistent [28].

**3.3. Relationships between motivation, grit, the choice of college major, and academic performance (Link c, g, h, j)**:
Recent advancements in neuroscience and neuroimaging technologies have opened up many latent insights regarding intrinsic motivation, growth mindset, and academic performance [29]. Specifically, the growth mindset has close ties with academic performance, engagement, and the ability to welcome new challenges within one's field [30, 31]. These papers have allowed us to infer the relationship between grit and academic performance (**link c**).

On the other hand, it has been shown that "*intrinsic learning motivation has a significant influence on choice of majors at state universities*" [32]. Combining these two findings, one can infer the association between the choice of major and intrinsic motivation (**links g, j**). Specifically, junior-standing students, who had decided their major(s) in their sophomore year of study, are expected to be intrinsically motivated and engaged to solve the challenges in their particular field. It is important to note that link h in the model has not been connected in [32] but there are studies in the literature suggesting that external factors such as salary, market demands have huge influence on college major choices [33] (e.g., **link h**).

**3.4. Our contributions and subsequent model interpretation:**
In this paper, our contributions are as follows:
  a) Relating brain fingerprint and cognitive diversity with grit (causal **link a**) and motivation (causal **link b**)
  b) Connecting cognitive diversity with intrinsic motivation and college major choice and performance (**Path: b-e-g-j**)
  c) Connecting cognitive diversity with extrinsic motivation and college major choice and performance (Path: **b-f-h-j**)
  d) Correlating grit with intrinsic motivation (**link d**)

The remaining piece of the puzzle is then to figure out how to incorporate intrinsic motivation in the context of interdisciplinary course curation and subsequent pedagogical practices.

**Section 4. Implication for teaching and practice**
Due to the natural complexity of real-world problems, most often, no challenge can be effectively solved within a short period of time. From an educational perspective, it is even harder to "simulate" the hardness of these challenges in a classroom context. A couple of useful questions that an engineering educator might address when considering students' neuro diversity:

a) How should an educator respond to the needs to train students to solve complex, multidisciplinary engineering problems in an academic setting?
b) What we do know thus far is the fact that assessment strategies such as quizzes would not be effective in measuring or reflecting students' level of tackling new challenges?

Referring back to Section 3, we also note that the thickness of *e* and *f* arrows indicates the effectiveness of a multidisciplinary engineering education system in which we would like to maximize the thickness of arrow e, ideally to match the associated grit level designed by the program. To maximize the engagement of interdisciplinary students, quizzes should be eliminated and replaced with a highly complex set of homework problems that span multiple disciplines. For instance, constructional engineers are to be encouraged or even required to solve issues on highway planning and bridge construction using stochastic models. The pedagogical innovation here is to leverage the theorized link of cognitive diversity and intrinsic motivation (Section 3.1). As such, students from different majors will find problems related to their chosen major of study interesting and intellectually stimulating. Cognitively, they would be motivated to spend a good amount of time understanding the context of the problem and trying to solve it. Another inferred solution based on the conceptual model in Section 3 is to drive students' learning through project-based [34], challenge-based [35], and experiential [36] learning. As much as project-based learning is fairly established among educators nowadays, not much of a relationship has been established between motivation and cognitive diversity in the context of project-based approach. Through our conceptual model, we would hypothesize that cognitive diversity drives grit and intrinsic motivation, which are key elements in project-based learning success.

## 5. Conclusion, limitation and future directions

In this work, we hypothesized neuro-induced linkages between cognitively different learners and its influence on the knowledge acquisition process. We apply this in the context of multidisciplinary engineering education and proposed possible pedagogical approaches that could be inferred from the conceptual framework. Our study would serve as a pioneer model to theorize key components of well-designed multidisciplinary engineering coursework which centers around the key question: how do we motivate different students of different backgrounds who are interested in many different problems in the same pedagogical deliverables settings. The most vital limitation that is worth pointing out is that our study lacks actual verification from the classroom. To continue strengthening the conceptual framework, we need to first verify the inferred pedagogical suggestions extracted from our conceptual framework by collecting students' opinion forms and feedback. Future studies should also include the effect of teamwork and diversity in the context of project-based learning which could include the evaluation of our model in the context of CATME [37].

Acknowledgements
Dr. Duy Duong-Tran would like to thank Trang Ngoc Nguyen, MBA for her contributions to the paper format.